\documentclass[aps,twocolumn,floats,superscriptaddress,prd,nofootinbib]{revtex4-1}
\usepackage{graphicx, epsfig, bm, amsmath}

\usepackage{color}
\usepackage{hyperref}
\usepackage{ wasysym }
\begin{document}


\newcommand{\lcdm}{$\Lambda$CDM}

\newcommand{\gpr}{G^{\prime}}

\newcommand{\fnl}{f_{\rm NL}}
\newcommand{\curv}{{\cal R}}

\definecolor{darkgreen}{cmyk}{0.85,0.2,1.00,0.2}
\newcommand{\ab}[1]{\textcolor{red}{[{\bf AB}: #1]}}
\newcommand{\wh}[1]{\textcolor{blue}{[{\bf WH}: #1]}}
\newcommand{\gB}{g_B}
\newcommand{\WP}{W}
\newcommand{\XP}{X}
\newcommand{\B}{B^{\rm Bulk}}
\newcommand{\tm}{m}
\newcommand{\sh}{M}

\newcommand{\aap}{Astron. Astrophys.}


\pagestyle{plain}

\title{Equivalence Principle Violation in Weakly Vainshtein-Screened Systems}

\author{Alexander V. Belikov}
\affiliation{Institut d'Astrophysique de Paris, UMR 7095 CNRS, Universit\'e Pierre et Marie Curie, 98 bis Boulevard Arago, Paris 75014, France}
        
\author{  Wayne Hu}
\affiliation{Kavli Institute for Cosmological Physics, Department of Astronomy \& Astrophysics, Enrico Fermi Institute, University of Chicago, Chicago, IL 60637}

\begin{abstract}
Massive gravity, galileon and braneworld models that modify gravity to explain cosmic acceleration utilize the nonlinear field interactions of the Vainshtein  mechanism to screen fifth forces in high density regimes.    These source-dependent interactions  cause apparent equivalence principle violations.  In 
the weakly-screened regime violations can be especially prominent since the fifth forces are at near full strength. Since they can also be calculated
perturbatively,   we derive analytic solutions for  illustrative cases:
 the motion of massive objects in compensated shells and voids and infall toward 
 halos that are spherically symmetric.   Using numerical techniques we show that these solutions are valid until the characteristic scale becomes comparable to the Vainshtein radius.    We find  a relative
 acceleration of more massive objects toward the center of a void 
 and a reduction of the infall acceleration that increases with the mass
 ratio of the halos which can in principle be used to test the Vainshtein screening mechanism.
\end{abstract}

\maketitle

\section{Introduction}
\label{sec:intro}

In models that seek to explain the current acceleration of the cosmic expansion by modifications to gravity or the addition of universal fifth forces on cosmological scales, such modifications must be hidden from local tests by so-called screening mechanisms.
Screening mechanisms invoke nonlinearity in the  equations  of motion for the field that mediates the extra force.   

The Vainshtein screening mechanism  \cite{Vainshtein:1972sx, Babichev:2010jd} was first introduced in the context of massive gravity to suppress the propagation of additional
helicity modes \cite{vanDam:1970vg,Zakharov:1970cc}. Here nonlinear derivative
interactions of the field act to screen the fifth force within the
so-called Vainshtein radius around a matter source. The Vainshtein
mechanism occurs not only in modern incarnations of Boulware-Deser
\cite{Boulware:1972zf} ghost-free massive gravity
\cite{deRham:2010kj, Chkareuli:2011te, Koyama:2011yg, Sbisa:2012zk} but also in galileon cosmology
\cite{Nicolis:2008in, Deffayet:2009wt, Burrage:2010rs, Kaloper:2011qc, DeFelice:2011th, Kimura:2011dc, deRham:2012fw} and braneworld models \cite{Deffayet:2001uk, Lue:2002sw, Lue:2004rj, Koyama:2007ih, Schmidt:2009sg,Schmidt:2009sv,Schmidt:2009yj}.

In these models,
all bodies accelerate equivalently in the total field of the fifth force 
and hence obey a 
microscopic equivalence principle.  Nonetheless
 screened bodies do not move as test bodies in an  external field leading
 to an apparent or macroscopic violation of the equivalence
 principle \cite{Hui:2009kc}.
Instead, nonlinearity in the field interactions causes interference between the external and body field  in forming the total field \cite{Hu:2009ua}.   
In the Vainshtein mechanism, this interference occurs when the second spatial derivatives of the fields becomes large enough that the self-interaction terms become important, i.e.\ within the Vainshtein radius of the external and body sources.

A general technique was recently introduced
for determining such effects in two-body systems by considering the effective density generated by the nonlinear interaction \cite{Hiramatsu:2012xj}.  It was applied to the Earth-Moon system, 
which exhibits strong screening since the orbit of the Moon is well within the Vainshtein radius of both the Earth and the Moon.  Violations of the equivalence principle thus appear as a small mass-dependent correction on the already suppressed effect of anomalous perihelion precession.   Here we study the weakly-screened limit where the Vainshtein mechanism is only beginning
to operate.   These cases have the advantage that  fifth forces are only weakly suppressed and that equivalence principle violations are  analytically solvable by a 
perturbative expansion.   Weak screening is applicable to cosmological situations such as 
voids and the outskirts of dark matter halos.

The outline of this paper is as follows.   We begin   in \S \ref{sec:weak} with a review of the Vainshtein
mechanism and develop the effective density approach for the weak-screening
regime.   In \S \ref{sec:analytic}, we consider several
examples where the acceleration of massive bodies can be analytically calculated.
In \S \ref{sec:beyond}, we test the limits of validity for the weak-screening 
approximation.
We discuss these results in \S \ref{sec:discussion}.

\section{Weak Screening}

\label{sec:weak}

For models that exhibit Vainshtein screening,  the ordinary Newtonian potential is modified by
the addition of a scalar field $\phi$ which itself obeys a nonlinear Poisson equation
\begin{equation}
 3\beta \nabla^2\phi +
N[\phi,\phi] = 8\pi G \delta\rho, \label{eq:phievo}
\end{equation}
where $\beta$ is a parameter that determines the coupling to matter density fluctuations from the  cosmological mean $\delta \rho = \rho -\bar \rho$.
For definiteness, we take the Dvali-Gabadadze-Porrati (DGP) braneworld example \cite{Dvali:2000hr}
 where the derivative operator is given by
\begin{equation}
N[\phi_1,\phi_2] = r_c^2 \left[ \nabla^2\phi_1 \nabla^2\phi_2 -
\nabla_i\nabla_j \phi_1 \nabla^i \nabla^j \phi_2 \right]
\end{equation}
in the quasistatic approximation.  Here $r_c$ is the crossover scale and determines the strength of the nonlinear interactions.
We have written out this derivative interaction term in a
general bilinear form to facilitate its use in two-body systems \cite{Hiramatsu:2012xj}.

For a single, spherically symmetric body of mass $M$ it is straightforward to show that 
fifth forces are screened within the Vainshtein scale (e.g. \cite{Schmidt:2009yj})
\begin{equation}
r_{*M}= \left(\frac{16 G M r_c^2}{9\beta^2}\right)^{1/3}.
\end{equation}
For example, we can define a weak-screening 
regime exterior to this scale where \cite{Hiramatsu:2012xj}
\begin{equation}
\phi_M \approx  -\frac{2 G M}{3 \beta r}\left( 1- \frac{1}{16}\frac{r_{*M}^3}{r^3} \right), \quad r \gg r_{*M}.
\label{eqn:singlefield}
\end{equation}

For two bodies, there is an additional screening that can be expressed as the interference between the 
fields of the individual bodies.
Given two  sources $\delta \rho_M$ and $\delta \rho_m$, which 
individually produce fields 
$\phi_M$ and $\phi_m$ satisfying
\begin{equation}
 3\beta \nabla^2\phi_{M,m} +
N[\phi_{M,m},\phi_{M,m}] = 8\pi G \delta\rho_{M,m},
\end{equation}
jointly $\phi_M+\phi_m$ no longer solves Eq.~(\ref{eq:phievo}) for $\delta\rho_M+\delta\rho_m$.
Instead, the interference field \cite{Hiramatsu:2012xj}
\begin{equation}
\phi_\Delta \equiv \phi -  \phi_M -  \phi_m,
\end{equation}
solves Eq.~(\ref{eq:phievo}) if
 \begin{eqnarray}
3 \beta \nabla^2 \phi_\Delta 
&=&  - 2N[\phi_M,\phi_m] -2 N[\phi_M+\phi_m,\phi_\Delta] 
\nonumber\\
&&
- N[\phi_\Delta,\phi_\Delta] .
\label{eqn:phidelta}
\end{eqnarray}
Thus the real density fields $\delta \rho_M$ and $\delta \rho_m$ are eliminated in favor of an effective density field given
by the nonlinear interaction terms.
Note that the real density fields can themselves be composite systems of $n$ individual bodies 
so long as the solutions $\phi_M$ and $\phi_m$ are known.   In general
this system remains a nonlinear Poisson equation, amenable 
only to numerical techniques \cite{Oyaizu:2008tb,Schmidt:2009sg,Chan:2009ew}.

Now let us consider the weak-screening limit.   By taking $r_c^2 \rightarrow 0$, we can 
iteratively solve for joint screening effects order-by-order in $r_c^2$.   To leading order
 $\phi_\Delta = {\cal O}(r_c^2)$ and Eq.~(\ref{eqn:phidelta}) can be approximated as
\begin{equation}
3\beta \nabla^2 \phi_\Delta^{(0)} = - 2N[\phi_M,\phi_m],
\label{eqn:phi0}
\end{equation}
which is now a linear Poisson equation with an external effective density.
Using this leading order expression, we can find the first order correction by solving
\begin{equation}
3 \beta \nabla^2 \phi_\Delta^{(1)} =-2 N[\phi_M+\phi_m,\phi_\Delta^{(0)}] ,
\label{eqn:phi1}
\end{equation}
etc.  Convergence of this series checks the validity of the approximation.   By dimensional analysis, the series should converge so long as $r_{*M}/s \ll 1$ and $r_{*m}/s \ll 1$ where $s$
is the typical physical scale of the system, e.g. the separation between bodies (see \S \ref{sec:beyond}).

The weak-screening limit for two-body interactions can  apply even if near the
individual bodies the individual fields enter a strong-screening regime.
For example near the location of mass $m$, $\phi_m$ may actually be strongly self-screened  if the physical size of the body is smaller than
 its Vainshtein radius.   All that is required for two-body weak screening
is that the interference source $N$ is dominated
by the  weakly-screened far-fields and that it generates an interference
field 
$\phi_\Delta$ that is nearly a pure gradient across the Vainshtein radius of
$m$  \cite{Hui:2009kc}.   Higher order contributions from Eq.~(\ref{eqn:phi0}) are therefore small
since the second derivatives of $\phi_\Delta$  are small where 
those of $\phi_m$ are large.

\begin{figure}[t]
\centerline{ \psfig{file=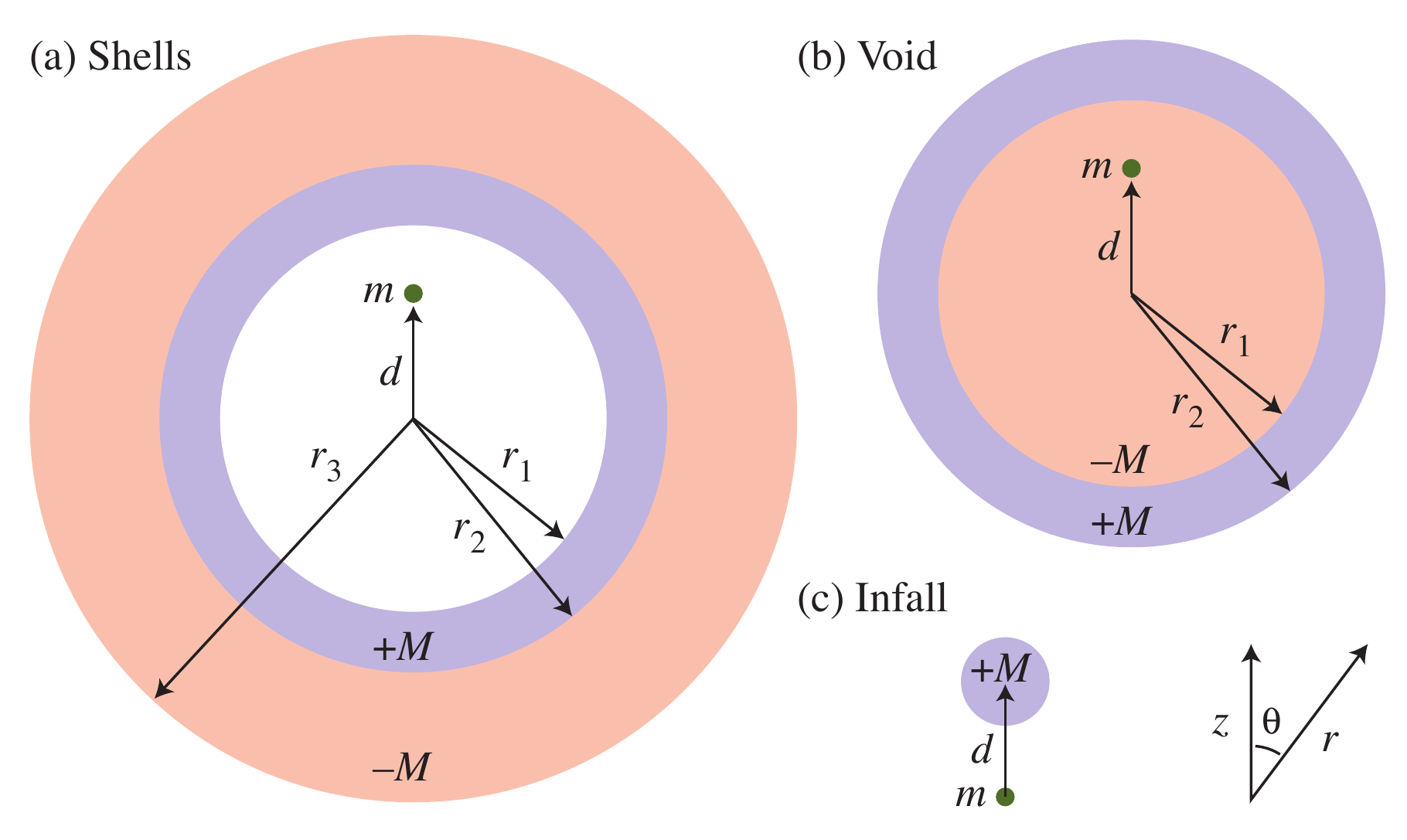, width=3.45in}}
\caption{Weak screening examples: (a) compensated shells with $\delta \rho_M=0$ in the interior and mass $\pm M$ in the shells; (b) compensated void with an underdensity of mass
$-M$ in the center and $+M$ in the shell;  (c) infall toward onto a spherical mass $M$.  The smaller point particle, whose motion we study, is labeled $m$.}
\label{fig:examples}
\end{figure}

\section{Analytic Examples}
\label{sec:analytic}

In this section, we consider three examples of equivalence principle violation in
weakly-screened systems where the mass-dependent acceleration of  bodies in the system can be solved analytically: empty, compensated mass shells; 
negative density 
fluctuation, compensated voids; and infall into a spherically symmetric body. 

\subsection{Shells}

 In Newtonian mechanics the shell theorem says that a body  inside
a spherically symmetric shell does not experience a force from the shell.   Despite the lack of a purely $1/r^2$ force under the Vainshtein mechanism
(see e.g.\ Eq.~\ref{eqn:singlefield}), it remains true that a test body of infinitesimal mass inside the shell does not 
experience a force (see e.g.~\cite{Schmidt:2009yj}).  However for a finite mass $m$,
 the nonlinear interaction between the body field and shell field causes a force
 unless the body is at the center of the shell.    This effect is common to
 gravitational field equations that are nonlinear.   In general relativity its impact
 is suppressed by the strength of the nonlinearity, i.e.\ the ratio of the Schwarzschild radius
 of the body to the separation $G m/s$, and in Modified Newtonian Dynamics it appears unsuppressed in the deeply nonlinear regime
 \cite{Dai:2008js}. 

To set up this first test case,
we take $\delta \rho_M$ to contain a constant-density shell that encloses a mass of $+M$ between $r_{1} < r \le r_{2}$.  In order to ensure compatibility with cosmological boundary conditions where the mean density fluctuation is zero, we take an additional outer shell of $-M$ between  $r_{2} < r\le r_{3}$.  Note that in the limit $r_3 \rightarrow \infty$ the result will be the same as an uncompensated single shell.
We obtain the $\phi_M$ solution to zeroth order in $r_c^2$  by superimposing
the Newtonian solutions for a tophat constant density enclosing a mass $M_{\rm th}$ within the radius $r_{\rm th}$
\begin{equation}
\phi_{\rm th} =  -\dfrac{2}{3\beta} \dfrac{G M_{\rm th}}{r_{\rm th}}  \begin{cases}
(\frac{3}{2} - \frac{r^2}{2r_{\rm th}^2} )
 & r\le  r_{\rm th} \\
 \frac{r_{\rm th}}{r} &  r > r_{\rm th}\\
 \end{cases}.
\end{equation}
Superimposing tophats of the appropriate mass, $\phi_M=$ const. for
$r\le  r_1$ and $0$ for $r>r_3$.    

\begin{figure}[t]
\centerline{\psfig{file=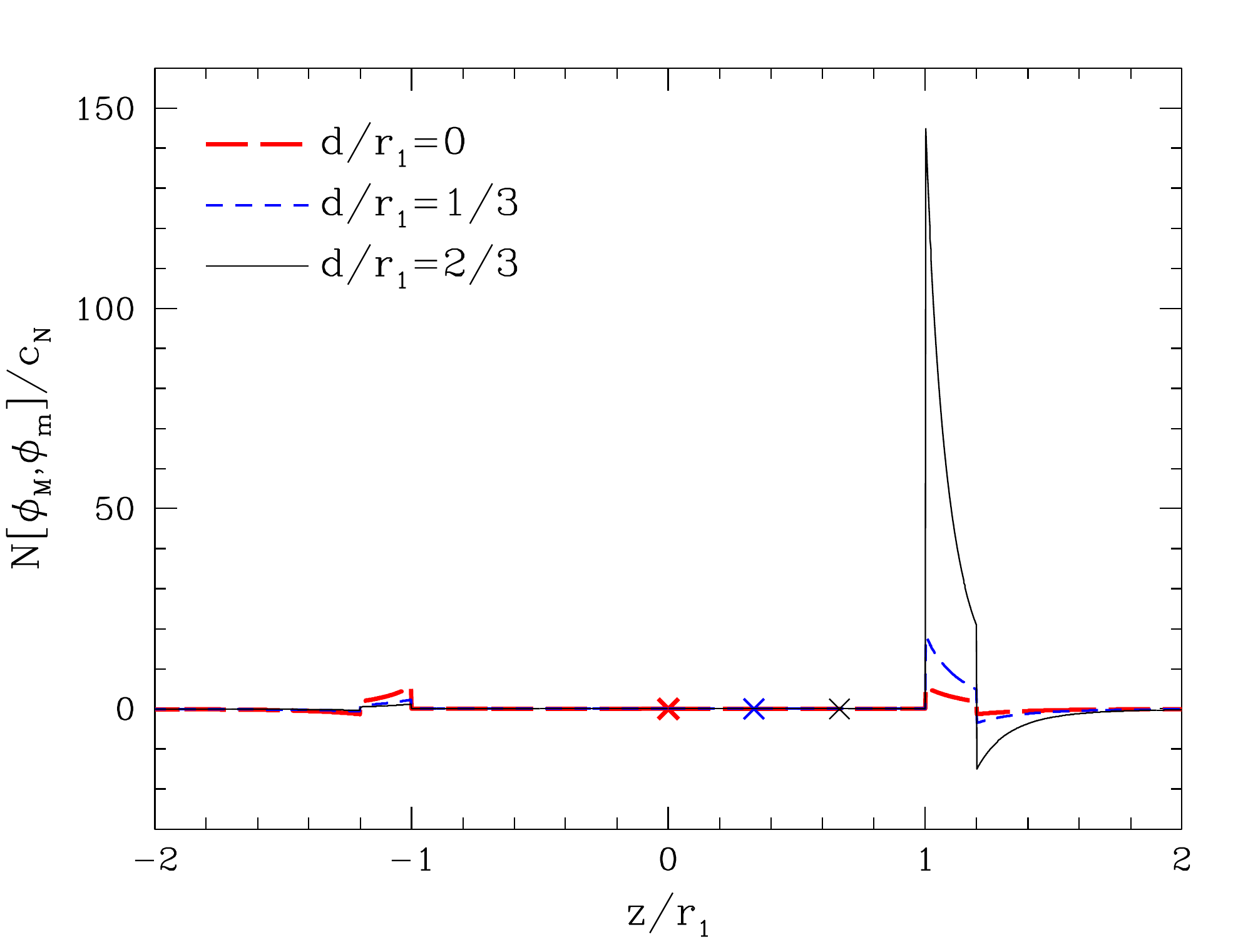, width=3.45in} }
\caption{Effective density as a function of $z= r\cos\theta$
 for the shells case of Fig.~\ref{fig:examples}a with $r_2=1.2 r_1$ and $r_3\rightarrow \infty$ 
and several choices of the displacement $d$ of the particle from the center of the shell (crosses).
As the particle is displaced  further along the $+z$ axis, the effective density in the
shells exhibits an increasing asymmetry which leads to the net force.}
\label{fig:N}
\end{figure}

Next
we take a small body of mass $m$ displaced along the $z$ axis at $z=d<r_1$ so as
to be inside the shell (see Fig.~\ref{fig:examples}a).  It suffices to consider this mass to be uncompensated.   If we place a compensating shell of
 $-m$ at $r>r_3$, we obtain identical results since there can be no interference
 with the  $\phi_M=0$ field there.
More specifically,
$N[\phi_{\sh},\phi_{\tm}]$ has compact support and
is only non-vanishing for $r_1 < r \le  r_3$
\begin{equation}
N[\phi_{\sh},\phi_{\tm}] =c_N r_1^6 F(r,\theta) \begin{cases}
\frac{r_1^3 }{ r_2^3-r_1^3} & r_1<r\le r_2 \\
 \frac{ r_3^3 }{ r_2^3-r_3^3}  & r_2<r\le r_3 \\
0 & {\rm else}\\
\end{cases}
\end{equation}
 with
\begin{equation}
F(r,\theta) = \frac{d^2+ 4 r^2 - 8 d r \cos\theta + 3 d^2\cos(2\theta)}{r^3(d^2+r^2 - 2 d r\cos\theta)^{5/2}}
\end{equation}
and
\begin{equation}
c_N =  \dfrac{2 r_c^2 G^2 M m}{3\beta^2 r_1^6} = \frac{3}{8} \frac{G M}{r_1^3} \left( \frac{r_{*m}}{r_1}
\right)^3,
\end{equation}
where $r_{*m}$ is the Vainshtein radius of $m$ in isolation.   

Even if $\phi_m$ were in the deep screening regime near the location of $m$, there would still be no effective density there since $\phi_M$
is constant inside the shell.  Thus we can safely take the limit of a point mass particle.
In Fig.~\ref{fig:N}, we show how the effective density changes as the point
mass is taken from the center of the shell toward the edge.  Note the asymmetry
that develops between the near and far side of the shell.

We  can build our solution for $\phi_\Delta$ from that of the following electrostatics-like Poisson equation
\begin{equation}
\nabla^2\Phi =
\begin{cases}
 0 & r \le r_0 
 \\
 -F(r,\theta) &  r_0 <  r< \infty 
 \\
 \end{cases} .
\end{equation}
More specifically, we are interested in calculating the potential
gradient $\nabla \Phi=\partial\Phi/\partial z$ 
at a position of the  particle $r=d$ and $\theta=0$, or the ``electric field" at that point
\begin{eqnarray}
\frac{\partial\Phi}{\partial z} &=& 
\frac{1}{2} \int_{r_0}^{\infty} r^2 d r  \int_0^\pi \sin\theta d\theta
 \frac{( r \cos\theta -d) F( r,\theta) }{( r^2+d^2-2  r d\cos \theta)^{3/2}}\nonumber\\
 &=& \frac{1}{d^5} I(r_0/d),
 \end{eqnarray}
 where
\begin{equation}
I(x) = \frac{x^2 +1}{4 x^2(x^2-1)^2} + \frac{1}{8 x^3} \ln 
\left( \frac{x-1}{x+1} \right).
\end{equation}
In the $d\rightarrow 0$ limit, $x\rightarrow \infty$ and 
$I(x) \approx 2/(3 x^6)$.

\begin{figure}[t]
\centerline{ \psfig{file=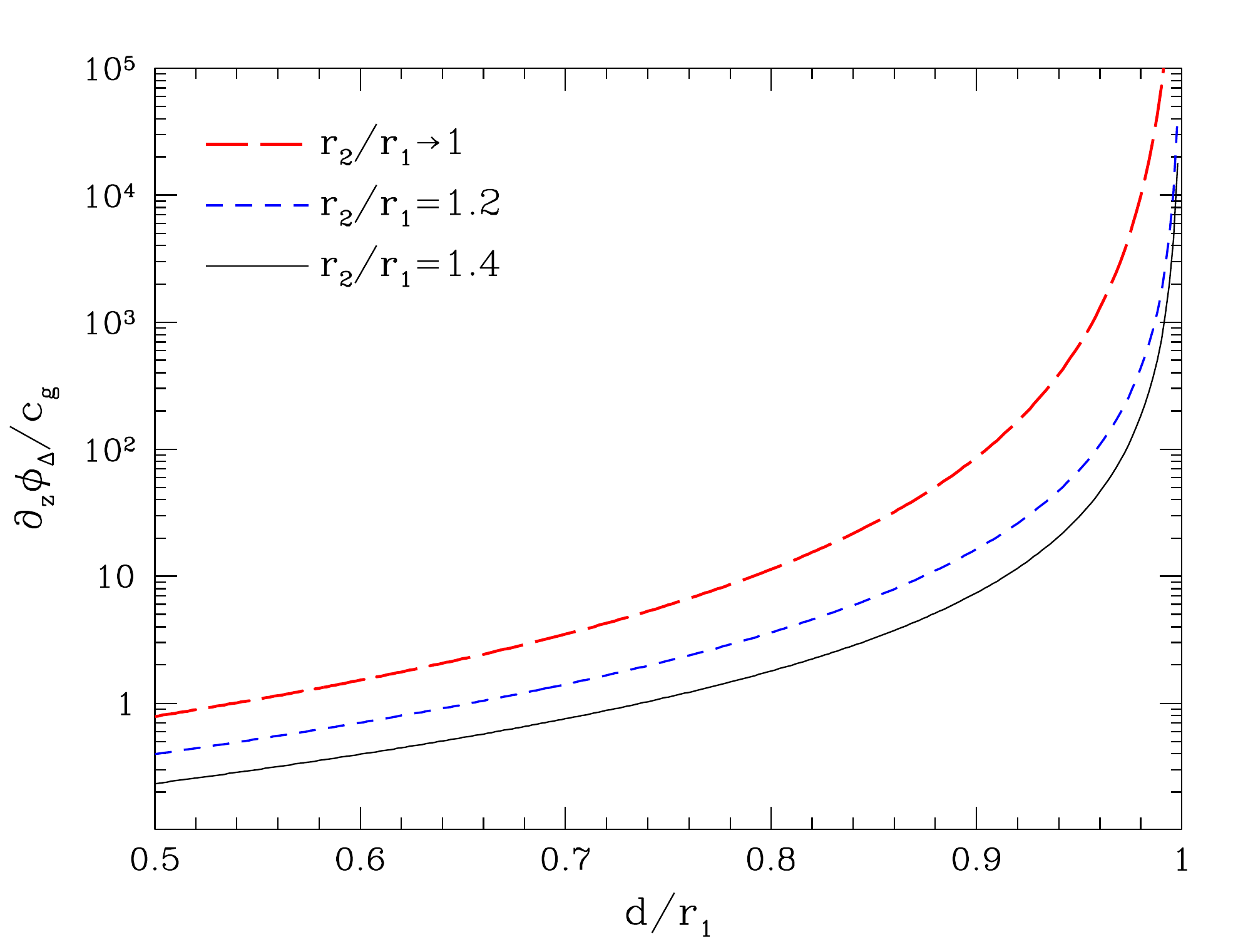, width=3.45in}}
\caption{Acceleration or field gradient  as a function of distance $d/r_1$ from center for various shell widths given by $r_2/r_1$ and $r_3 \rightarrow \infty$.  
As the 
particle approaches the shell, the acceleration increases as $s^{-3}$, where
$s= r_1-d$, with a proportionality constant that reaches a maximum 
as the shell width goes to zero $r_2/r_1 \rightarrow 1$.}
\label{fig:acceleration}
\end{figure}

We again superimpose these solutions with the appropriate $r_0$ to construct
the full solution
\begin{eqnarray}
\frac{\partial\phi_\Delta}{\partial z} \Big|_{m} &=& c_g \left( \frac{r_1}{d} \right)^5
 \Big\{ \frac{r_1^3}{r_2^3-r_1^3} [I(r_1/d)-I(r_2/d) ]\nonumber\\
 &&
 -\frac{r_3^3}{r_3^3-r_2^3} [ I(r_2/d)-I(r_3/d) ]
\Big\} ,
\label{eqn:shellacc}
   \end{eqnarray}
  where
  \begin{equation}
  c_g =  \frac{4r_c^2}{9\beta^3} \frac{G^2 M m}{r_1^5}= \frac{1}{4\beta} \frac{GM}{r_1^2} 
\left(  \frac{r_{*m}}{r_1} \right)^3.
 \end{equation}Eugeny Babichev
 Note that $c_g$ scales as the shell acceleration 
 $G M/\beta r_1^2$ suppressed by the cube of the Vainshtein scale over the characteristic distance.

   Examples of the acceleration as a function of $d/r_1$ are shown in 
   Fig.~\ref{fig:acceleration}.
   One interesting limit is an infinitesimally thin shell $r_2\rightarrow r_1$ with vanishing impact from cosmological compensation
   $r_3 \rightarrow \infty$
   \begin{eqnarray}
\frac{\partial\phi_\Delta}{\partial z} \Big|_{m} &=&  c_g 
\left[ \frac{2}{3} \frac{d/r_1}{(1-d^2/r_1^2)^3 }\right].
 \end{eqnarray}
 Consider the cases where the particle is near the center
 or near the shell.   For the former case, 
 there is a suppression of $(r_{*m}/r_1)^3 (d/r_1)$ from the scale of the Newtonian acceleration of
 the shell mass.  The first factor accounts for the weak nonlinearity of the system.
 The second factor accounts for the fact that by symmetry the force must vanish
 if $d=0$.  
 
  
   In the opposite regime where the particle is closer to the shell than the center, 
   the acceleration scales strongly with  distance to shell $s= r_1-d$. In particular
   as the particle approaches the shell
     \begin{equation}
\lim_{s\rightarrow 0} \frac{\partial\phi_\Delta}{\partial z} \Big|_{m} = \frac{1}{6\beta}\frac{G M}{r_1^2}
 \left( \frac{r_{*m}}{2 s} \right)^3 ,
 \end{equation}
such that there is a suppression factor of the Vainshtein radius compared to the distance to the shell (see Fig.~\ref{fig:acceleration}).   

 In this case, we see the impact
of the nonlinearity of the field equations unobscured by symmetry.  Once the Vainshtein screened
regime of $m$ starts to overlap the shell, the field of the shell effectively
creates a boundary condition there, much like the establishment of a mirror
charge by a conductor.   Since the Vainshtein radius of the particle depends on
its mass, the equivalence principle is macroscopically broken: for an 
attractive fifth force, the acceleration of a particle towards the center will grow linearly with the mass $m$.
The sign of the effect is determined by the fact that the 
 Vainshtein correction in Eq.~(\ref{eqn:singlefield})
 makes the force fall away with distance less rapidly than $1/r^2$ and so the far side of the shell that contributes more force than the near side.

\subsection{Void}

A closely related and more observationally relevant case is a cosmological
void.   In this case instead of an interior at the mean cosmological density
and $\delta \rho_M =0$, we have an underdensity.   We can idealize the void
as a spherical tophat of spatially constant underdensity $\delta\rho_M = \bar\rho \Delta_V$ where 
$-1 \le \Delta_V<0$ for $r\le r_{1}$ with a total negative mass fluctuation of
\begin{equation}
-M  =  \frac{4\pi}{3} r_{1}^3 \bar \rho 
\Delta_V. \end{equation}
The void is surrounded by a positive density shell of mass $+M$ for $r_1 < r \le r_2$ such that the total mass fluctuation exterior to $r_2$ is zero (see Fig.~\ref{fig:examples}b).
The Vainshtein scale of such a void is
\begin{equation}
r_{*M} = r_1 \left( \frac{8\Omega_m H_0^2 r_c^2}{9\beta^2} |\Delta_V| \right)^{1/3}
\end{equation}
and so for $H_0 r_c/\beta \lesssim 1$, a void of any size $r_1$ is  in the weak-screening regime throughout its interior.

Despite the fact that the interior now has a finite $\delta \rho_M$, the void and shell
case are nearly identical.   The reason is that if 
$\phi_M$ is the $r^2$-field associated with a constant density source, then
\begin{equation}
N[\phi_M,\phi_m] = \frac{2}{3} r_c^2 \nabla^2\phi_M \nabla^2\phi_m
\end{equation}
regardless of the form of $\phi_m$ \cite{Hiramatsu:2012xj}.

For a particle of mass $m$, the effective density therefore vanishes inside the void except at the position of the test mass.    If we take the particle to be
a constant density spherical tophat with $\rho_m$ that vanishes outside its radius
$r_m \ll r_{1}$
\begin{equation}
N[\phi_M,\phi_m] 
=\frac{2r_c^2 G^2}{3\beta^2}   
\begin{cases}
\left( \frac{8\pi}{3}\right)^2
\bar\rho\Delta_V \rho_{m}  & r \le r_{1} \\
\frac{ r_{2}^3}{r_{2}^3-r_{1}^3} M m F(r,\theta) & r_{1} < r  \le r_{2}  \\
 0 & r>r_{2}
\end{cases} .
\end{equation}

The $r\le r_{1}$ term provides no acceleration at the position of the test mass by 
symmetry and the shell term gives
\begin{equation}
\frac{\partial\phi_\Delta}{\partial z} \Big|_{m} = 
c_g \left( \frac{r_1}{d} \right)^5
 \frac{r_{2}^3}{r_{2}^3-r_{1}^3} \left[I\left(\frac{r_1}{d}\right)-I\left(\frac{r_2}{d}\right) \right],
  \end{equation}
which has a form very similar to 
Eq.~(\ref{eqn:shellacc}). In particular the infinitesimal width $r_2\rightarrow r_1$
case has the  limiting behaviors
\begin{eqnarray}
\lim_{d/r_1\rightarrow 0}  \frac{\partial\phi_\Delta}{\partial z} \Big|_{m} &=& 
  \frac{1}{3\beta}\frac{G M}{r_1^2}
 \left( \frac{r_{*m}}{r_1} \right)^3 \frac{ d}{r_1}, \nonumber\\
 \lim_{s/r_1\rightarrow 0}  \frac{\partial\phi_\Delta}{\partial z} \Big|_{m} &=&
   \frac{1}{6\beta}\frac{G M}{r_1^2}
 \left( \frac{r_{*m}}{r_1} \right)^3 \left( \frac{r_{*m}}{2 s} \right)^3 ,
  \end{eqnarray}
  where recall $s=r_1-d$ is the separation between the particle and the shell.
  In the later limit, the shell and void cases are identical as one might expect.
  
Halos in compensated spherical voids therefore accelerate differently depending on their mass when they are near a Vainshtein radius $r_{*m}$ of the shell.
The Vainshtein radius of a halo scales with its virial radius.
Taking the virial radius of the halo $r_{H}$ as the radius out to which the average interior density reaches $\Delta_H \sim 200$, we obtain the Vainshtein radius of a halo of mass $m$
\begin{equation}
r_{*m} = r_H \left( \frac{8\Omega_m H_0^2 r_c^2}{9\beta^2} |\Delta_H| \right)^{1/3}.
\end{equation}
For  $H_0 r_c/\beta \lesssim 1$, halos that are near a virial radius of the 
compensating shell will be accelerated away from the shell.

\subsection{Infall}

Lastly we consider the problem of two point masses, or tophat constant
density bodies of radius negligible compared with their separation.  This problem
was solved numerically in the strong-screening regime in Ref.~\cite{Hiramatsu:2012xj}.  Here we consider the weak-screening regime as might be appropriate
for a satellite dark matter halo of mass $m$
 falling through the virial radius of a  parent halo of mass $M$.

In this case, each mass in isolation carries a field given by Eq.~(\ref{eqn:singlefield}), e.g. 
\begin{equation}
\phi_{m} \approx -\frac{2 G m}{3\beta R} ,
\end{equation}
for the satellite mass and likewise for the parent mass $M$.
 For convenience, we place $M$ at $(r=0,z=d)$ and $m$ at the origin $(r=0,z=0)$ (see Fig.~\ref{fig:examples}c).   The effective density
  becomes
\begin{equation}
 N[\phi_M,\phi_m] = 
 -\frac{2}{3} \frac{r_c^2}{\beta^2} G^2 M m F(r,\theta).
\end{equation}
The main difference between this case and the previous cases is that 
the effective density does not
have compact support and indeed diverges at the position of the two bodies.

Nonetheless, just like for the divergent physical densities of point masses,
we can still evaluate the forces induced by one body on the other.
At the position of the satellite mass $m$
\begin{eqnarray}
\frac{\partial \phi_\Delta}{dz} \Big|_{m} &=& 
 -\frac{2}{9} \frac{r_c^2}{\beta^3} G^2 M m 
  \int_0^\infty r^2 dr \int_0^\pi \sin\theta d\theta F(r,\theta) \frac{\cos\theta}{r^2} \nonumber \\
 &=& \frac{8}{27} \frac{r_c^2}{\beta^3} \frac{G^2 M m }{d^5} = \frac{1}{6\beta}\frac{G M r_{*m}^3}{d^5} .
\end{eqnarray}
Since $\phi_m$ is symmetric around the origin, it provides no acceleration there
and the total is
\begin{equation}
\left[ 
\frac{\partial \phi_M}{\partial z} +
\frac{\partial \phi_\Delta}{\partial z} \right]_{m}
= -\frac{2 GM }{3\beta d^2} \left( 1 -\frac{r_{*M}^3+r_{*m}^3}{4 d^3}  \right)
\end{equation}
such that the acceleration of $m$ from the field of $M$ is reduced by $(r_{*M}^3+r_{*m}^3)/d^3$.    Note that we have kept here the ${\cal O}(r_c^2)$ correction to $\phi_A$ from Eq.~(\ref{eqn:singlefield}) to maintain the same order in $r_c$ throughout.
The impact of the two-body interference for $m\ll M$ is thus muted by the 
fact that the main suppression still comes from the screening of the larger mass.
As the two masses become comparable, the interference can make a more
substantial reduction of the 
attractive force.

This calculation also illustrates the role of the effective density in restoring
momentum conservation or Newton's third law.  
By symmetry, this calculation also gives the acceleration of body $M$ from the field of $m$ 
\begin{equation}
\left[ 
\frac{\partial \phi_m}{\partial z} +
\frac{\partial \phi_\Delta}{\partial z} \right]_{M} = \frac{2 Gm }{3\beta d^2} \left( 1 -\frac{r_{*m}^3+r_{*M}^3}{4 d^3} \right).
\end{equation}
The sum of the two forces
\begin{equation}
m \left[ 
\frac{\partial \phi_M}{\partial z} +
\frac{\partial \phi_\Delta}{\partial z} \right]_{m} + M
\left[ 
\frac{\partial \phi_m}{\partial z} +
\frac{\partial \phi_\Delta}{\partial z} \right]_{M} =0,
\end{equation}
whereas without the interference the balance would be grossly violated for
$m/M \ll 1$ \cite{Hiramatsu:2012xj}.

\begin{figure}[t]
\centerline{ \psfig{file=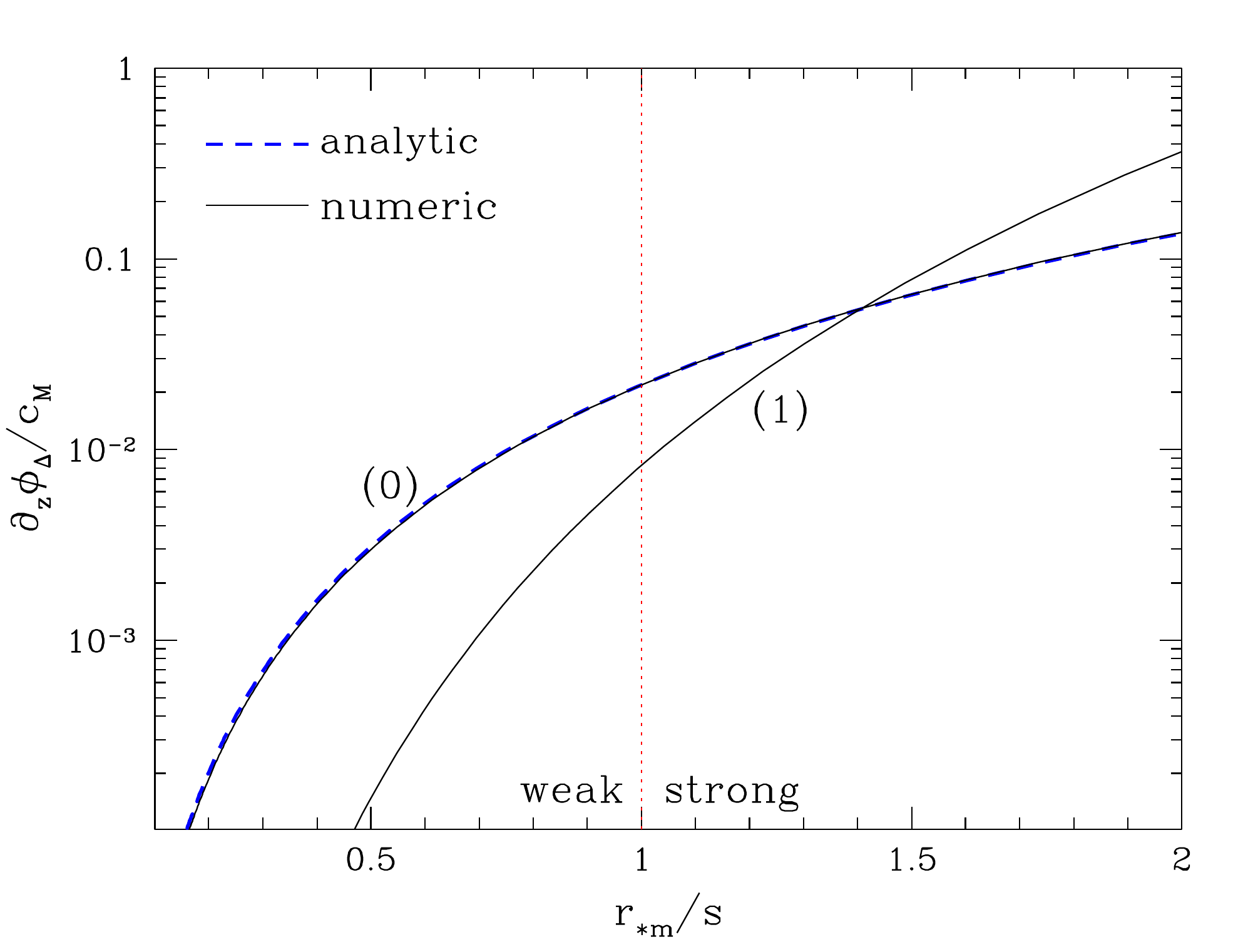, width=3.45in}}
\caption{Numerical test of weak screening.    Numerical results for the leading order acceleration ``(0)"  in the weak-screening approximation agree well with the analytic expression from Eq.~(\ref{eqn:shellacc}) while those for the first order correction ``(1)" from Eq.~(\ref{eqn:phi1}) 
indicate that the approximation is valid until the distance to the shell $s$ is smaller than a 
Vainshtein radius of the body $r_{*m}$ (vertical dashed line).  These terms have opposite sign, implying that the acceleration ceases to grow as rapidly in the strong screening regime.  See text for the shell parameters of this example.}
\label{fig:numerical}
\end{figure}

\section{Beyond Weak Screening}
\label{sec:beyond}

We expect the weak-screening regime to extend to the point at which the
Vainshtein radii of the individual bodies is comparable to the characteristic distances
in the system.
In our perturbative approach, the breakdown of weak screening is monitored
by the next to leading order term $\phi_\Delta^{(1)}$ in Eq.~(\ref{eqn:phi1}).
In this section we shall check to see when this breakdown occurs 
numerically.

For illustration purposes, we choose the shell test case (see Fig.~\ref{fig:examples}).
Given that the effective density $N$ vanishes for $r>r_3$, this case 
has periodic boundary conditions and can be solved efficiently with
fast Fourier transform techniques.  

  As an example, we take the outer
shell to be inscribed inside the cube of side length $L=1024$ pixels: $r_3=L/2$.  We take the
innermost  shell to be of radius $r_1= 3L/8$ and width $r_2-r_1=r_1/96$.  
We set the mass scales to be such that the Vainshtein scale of the
shell $r_{*M}=r_1/12$ and that of the mass $r_{*m}=r_1/24$.  Finally, we replace the
point mass $m$ with a constant density sphere of radius $r_m=r_1/96$.
In Fig.~\ref{fig:numerical}, we show the numerical results for the leading order 
acceleration and first order correction given by Eq.~(\ref{eqn:phi0}) and (\ref{eqn:phi1}) respectively.   Here we plot the acceleration in units  of the maximum Newtonian acceleration for
a test particle just outside of 
an infinitesimal shell of radius $r_1$
\begin{equation}
c_M = \frac{2 G M}{3\beta r_1^2}.
\end{equation}
The numerical calculation of the leading order $\phi_\Delta^{(0)}$ term matches
 the analytic expression, Eq.~(\ref{eqn:shellacc}) to excellent approximation showing the discretization onto the grid does not cause appreciable errors.
The first order correction $\phi_\Delta^{(1)}$  becomes comparable to the zeroth order
effect when $r_{*m}/s \sim 1$ implying that the weak-screening approximation
applies all the way to separations of a Vainshtein radius of $m$.  At this point the acceleration, 
which would be zero in the absence of nonlinearity, becomes a non-negligible fraction of the maximal
acceleration $c_M$. 

Note that for $M \gg m$, the validity of the weak-screening approximation extends to 
separations much smaller than $r_{*M}$ since nonlinear corrections for $M$ come in through
the characteristic scale of the shell
$r_{*M}/r_1$ rather than the characteristic scale of the separation $r_{*M}/s$.  Thus in the shell case, the qualitative rule that weak screening applies until the separation becomes comparable to the
Vainshtein radius of the individual sources is too conservative if applied to both
$r_{*m}$ and $r_{*M}$.

\section{Discussion}
\label{sec:discussion}

The weak-screening regime, where the Vainshtein mechanism is just beginning to suppress
fifth forces,  exhibits apparent or macroscopic violation of the equivalence
principle.  We have illustrated this effect through analytic calculations
of the acceleration of particles within a mass shell, compensated void, and
toward a spherically symmetric mass.  Numerical tests show that weak-screening 
approximation is valid until the bodies are separated by less than a Vainshtein radius.

With an attractive fifth force, 
massive objects such as dark matter halos in a mass shell or compensated void
are attracted to the center of the void with an acceleration proportional to their
mass.  A cosmological void is also naturally in the weak-screening regime
for the cosmologically interesting case where $H_0 r_c/\beta \sim 1$.   Since the Vainshtein
radius of a halo is comparable to its virial radius, the maximal effect will be on halos that
are close to a virial radius of the edge.   
  This effect has potential observable consequences for
mass segregation near the edge of voids, but it remains to be seen in cosmological
simulations whether deviations from the idealizations of spherical symmetry, perfect compensation and 
constant underdensity mask this effect.  

For the infall problem, the two-body interference predicts a reduction of
major mass mergers where the bodies are of comparable mass.   The infall
case also shows how the interference restores Newton's third law or
momentum conservation in the joint system.

These effects are signatures of the Vainshtein mechanism, which is itself
common to massive gravity, galileon and braneworld scenarios.
 Our analytic calculations serve as simple illustrations that expose new aspects
 of the mechanism though constructing realistic
cosmological tests  of it will require going beyond the idealizations considered here.

\acknowledgements
We thank T. Hiramatsu, E. Jennings, K. Koyama, Y. Li, F. Schmidt and E. Babichev for useful conversations.
AB was supported in part by ERC project 267117 (DARK) hosted by  Universite Pierre et  Marie Curie - Paris 6 and acknowledges the hospitality of KICP where part of this work was completed.
WH was supported  by the KICP  through grants NSF PHY-0114422 and NSF PHY-0551142 and an endowment from the Kavli Foundation and its founder Fred Kavli, U.S.~Dept.\ of Energy contract DE-FG02-90ER-40560 and the David and Lucile Packard Foundation.  WH acknowledges the hospitality of IAP where part of this work was completed.  

\vfill

\clearpage

\bibliography{BelHu12.bib}

\end{document}